\documentclass[preprint,prc]{revtex4}
\usepackage{graphicx}
\def\beq{\begin{equation}}
\def\eeq{\end{equation}}
\def\beqa{\begin{eqnarray}}
\def\eeqa{\end{eqnarray}}
\def\MeV{\nobreak\,\mbox{MeV}}
\def\GeV{\nobreak\,\mbox{GeV}}

\def\pli{p^\prime}

\usepackage{color}

\usepackage{ulem}

\begin{document}
\title{$B_s B^* K $ and $B_s B K^*$   vertices using QCD sum rules}
\author{A. Cerqueira Jr.,  B. Os\'orio Rodrigues}
\affiliation{Instituto de F\'{\i}sica, Universidade do Estado do Rio de
Janeiro, Rua S\~ao Francisco Xavier 524, 20550-900, Rio de Janeiro, RJ, Brazil. }
\author{M. E. Bracco}
\affiliation{Faculdade de Tecnologia, Universidade do Estado do Rio de Janeiro,
Rod. Presidente Dutra Km 298, P\'olo Industrial, 27537-000 , Resende, RJ, Brazil.}
\author{M. Nielsen}
\affiliation{Instituto de F\'{\i}sica, Universidade de S\~{a}o Paulo,
  C.P. 66318, 05389-970 S\~{a}o Paulo, SP, Brazil}

\begin{abstract}
The form factors  and the coupling constant of the
$B_s B^* K $ and $B_s B K^*$  vertices are calculated using the QCD sum rules method.
Three point correlation functions are computed considering both the heavy and light
mesons off-shell in each vertex, from which, after an extrapolation of the QCDSR results at the pole
of the off-shell mesons, we obtain the coupling constant of the vertex.
The form factors obtained have different 
behaviors but their simultaneous extrapolation reach the same value of the
coupling constant $g_{B_s B^* K}=8.41 \pm 1.23 $ and $g_{B_s BK^*}=3.3 \pm 0.5$.
 We compare our result with
other theoretical estimates and compute the uncertainties of the method.
\end{abstract}

%\pacs{14.40.Lb,14.40.Nd,12.38.Lg,11.55.Hx}

\maketitle

\section{Introduction}
In the recent years, many new charmonium and bottomonium states have been observed
at the B-factories. As an example, in the bottomonium sector, the Belle
Collaboration reported the observation of two
charged narrow structures in the $\pi^\pm\Upsilon(nS)~(n=1,2,3)$ and
$\pi^\pm h_b(mP)~(m=1,2)$ mass spectra of the $\Upsilon(5S)\to\Upsilon
(nS)\pi^\pm$ and $\Upsilon(5S)\to h_b(mP)\pi^\pm$ decay processes
\cite{bellezb}. These narrow structures were called $Z_b(10610)$ and
$Z_b(10650)$. As pointed out by the Belle Collaboration, the proximity
of the $B\bar{B}^*$ and $B^*\bar{B}^*$ thresholds and the $Z_b(10610)$ and
$Z_b(10650)$ masses suggests that these states could be interpreted as weakly
bound $B\bar{B}^*$ and $B^*\bar{B}^*$ states. In particular, using the
one-boson exchange model and considering $S$-wave and $D$-wave mixing, the
authors of Ref.~\cite{liu1} were able to explain both, $Z_b(10610)$ and
$Z_b(10650)$, as $B\bar{B}^*$ and $B^*\bar{B}^*$ molecular states.
The main
ingredients in the one-boson exchange model are the effective Lagrangians,
that describe the strong interactions between the heavy and light mesons. These
Lagrangians are characterized by the strong coupling constants in the considered
vertices which, in general, are not known. These heavy-heavy-light mesons coupling
constants are fundamental objects, since they can provide essential information
on the low energy behavior of the QCD. Depending on their numerical values,
a particular molecular state may or may not be bound. Therefore, it is really
important to have reliable ways to extract these values based on QCD
calculations. However, such low-energy hadron interaction lie in a region which is
very far away from the perturbative regime. Therefore, we need some
non-perturbative approaches, such as the QCD sum rules (QCDSR) \cite{svz,rry,SNB}, to
calculate the form factors and coupling constants of these vertices. There are already
some QCDSR calculations for the heavy-heavy-light vertices like the $B^*B\pi$
\cite{Navarra:2000ji}, $B_{s0}BK$ \cite{Bracco:2010bf}, $B_s^*BK^*$ \cite{Azizi:2010jj},
 $B^*B^*\rho$ \cite{Cui:2012wk}, $B_sBK_0^*$, $B_s^*BK_1$ \cite{Sundu:2011vz} and $B_s^* B K$ \cite{CerqueiraJr2012130}.
In the charm sector, various vertices were evaluated with this approach and the results
are systematized in \cite{Bracco:2011pg}.
Here we calculate the form factor and the coupling constant at the $B_sB^*K$ and $B_sBK^*$
vertices in the framework of three-point QCDSR. More specifically, we evaluate the
$g_{B_sB^*K}(Q^2)$ and $g_{B_sBK^*}(Q^2)$ form factors in three different ways, considering, 
one by one, each one of the mesons in the vertex to be off-shell. From these form factors, 
we extract the $g_{B_sB^*K}$ and the $g_{B_sBK^*}$ coupling constants.

\section{The QCD sum rule for the  $ B_s B^* K$ and $B_s B K^*$ vertices}

To perform the QCDSR calculation and obtain the form factors and coupling constants of the 
vertices $B_s B^* K$ and $B_s B K^*$, we follow our previous works, as 
Ref.~\cite{Bracco:2011pg}. 
The starting point is the three-point correlation function given by :
\begin{equation}
 \Pi_{\mu(\nu)}^{(B_s)}(p, p^{\prime})=\int d^4 x d^4 y
\left<0\right|T\{{j^{K^{(*)}}_{\mu}}(x){j^{B_s}}^{\dagger}(y){j_{(\nu)}^{B^{(*)}}}^\dagger(0)
\}\left|0\right> e^{ip^{\prime}\cdot x}e^{-iq\cdot y},
\label{correBsoff}
\end{equation}
for the $B_s$ meson off-shell, and:
\begin{equation}
 \Pi^{(K)}(p, p^{\prime})_{(\mu)\nu}=\int d^4 x d^4 y
\left<0\right|T\{{j^{B^{(*)}}_{(\mu)}}(x){j^{K^{(*)}}_{\nu}}^{\dagger}(y){j^{B_s}}^\dagger(0)
\}\left|0\right> e^{ip^{\prime}\cdot x}e^{-iq\cdot y},
\label{correkoff}
\end{equation}
for the $K$ or $K^*$ meson off-shell. In Eqs.~(\ref{correBsoff}) and (\ref{correkoff}), $q= p - 
p^{\prime}$ is the momentum of the off-shell meson and $p$ and $p^{\prime}$ are the momentum 
of other ones. The currents $j^{(M)} $ are the currents associated with each meson in the 
vertex and contain the quantum information about the state. 
The correlation functions in Eqs.~(\ref{correBsoff}) and (\ref{correkoff}) allow to obtain 
two different form factors corresponding to the same vertex. In this way, the vertex is 
tested by two different mesons, the heavier and the lighter ones in the corresponding vertex. 
The calculation of these two correlation functions allows to reduce the uncertainties of the 
evaluation of the coupling constant of the vertex~\cite{Bracco:2011pg}.

Equations (\ref{correBsoff}) and (\ref{correkoff}) contain different numbers of
Lorentz structures, and for each structure, we can write a different sum rule. In principle 
all the structures would give the same result. However, due to different approximations
each structure can lead to different results. Therefore, one has to choose the structures
less sensitive to the different approximations.  To obtain the sum rule, these  functions are 
calculated in two different ways: using quarks degrees of freedom -- the QCD 
side; and  using hadronic degrees of freedom -- the phenomenological side. In the QCD side, the 
correlators are evaluated using Wilson's operator product expansion 
(OPE). The duality principle allows us to obtain an interval in which both representations 
are equivalent. Therefore, in this region, we can obtain the QCD sum rule from where the form 
factors are evaluated. To improve the matching between the two sides, we perform a Borel 
transformation to both QCD and phenomenological sides.

\subsection{The QCD side}

The QCD side is obtained using the following meson currents for the $B_s B^* K $ vertex:
\beqa
&&j^{B^*}_{\mu}(x)= \bar b \gamma_{\mu} q,  \nonumber \\
&&j^{B_s}_5(0)= i \bar b  \gamma_5 s, \nonumber \\
&&{j_{\nu}^K}(y)=  \bar s \gamma_{\nu} \gamma_5 q \eeqa
and the following ones for the  $B_s B K^*$ vertex:
\beqa
&&j^{B}_5(x)= i \bar q \gamma_5 b,  \nonumber \\
&&j^{B_s}_5(0)= i \bar b  \gamma_5 s,  \\
&&{j^{K^*}_{\mu}}(y)=  \bar q \gamma_{\mu} s.  \nonumber \eeqa
Here, $q$, $s$ and $b$ are the light, strange and bottom quark fields 
respectively. Each one of these currents has the  quantum numbers of the associated meson. 
In the case of $K$ off-shell meson, we use, as usual in QCDSR, the pseudo scalar current for 
it, see refs. $D^*D\pi$ \cite{Navarra:2000ji}, $D^*_s D K $ \cite{angelo06} and $B_s^* B K$ \cite{CerqueiraJr2012130}.
%\beqa
%&&{j^K_{5}}(x)= i \bar q \gamma_5 s.
%\eeqa
The general expression for the vertices has different structures, which can be written in 
terms of a double dispersion relation over the virtualities $p^2$ and ${\pli}^2$, holding 
$Q^2= -q^2$ fixed:

\begin{equation}
\Gamma(p^2,{\pli}^2,Q^2)=-\frac{1}{4 \pi^2}\int_{s_{min}}^\infty ds
\int_{u_{min}}^\infty du \:\frac{\rho(s,u,Q^2)}{(s-p^2)(u-{\pli}^2)}\;,
\;\;\;\;\;\; \label{dis}
\end{equation}
where the spectral density $\rho(s,u,Q^2)$ can be obtained from the Cutkosky's rules.

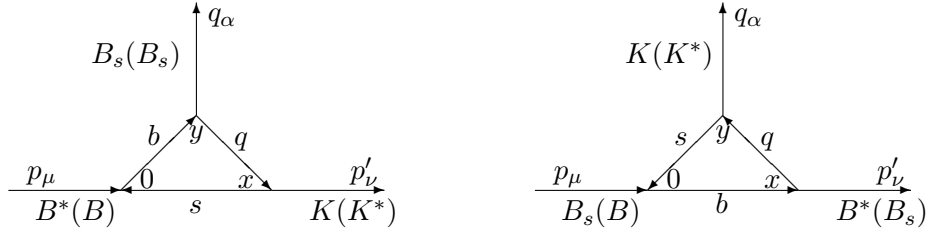
\begin{figure}[ht]

\begin{picture}(12,3.5)
% Left diagram
\put(0.0,0.5){\vector(1,0){1.5}}
\put(3.5,0.5){\vector(-1,0){2}}
\put(3.5,0.5){\vector(1,0){1.5}}
\put(1.5,0.5){\vector(1,1){1}}
\put(2.5,1.5){\vector(1,-1){1}}
\put(2.5,1.5){\vector(0,1){1.5}}
%\put(2.5,3){\vector(0,-1){1.5}}
\put(2.65,2.75){$q_\alpha$}
\put(0.25,0.65){$p_\mu$}
\put(4.55,0.65){$p'_\nu$}
\put(2.4,0.2){$ s$}
\put(1.85,1.1){$b$}
\put(3,1.1){$q$}
\put(2.4,1.2){$y$}
\put(1.75,0.53){$0$}
\put(3.05,0.53){$x$}
\put(1.1,2.2){$B_s (B_s)$}
\put(0.35,0.1){${B^{*} (B)}$}
\put(4,0.1){{$K (K^*)$}}
% Right diagram
\put(7,0.5){\vector(1,0){1.5}}
%\put(10.5,0.5){\vector(-1,0){2}}
\put(8.5,0.5){\vector(1,0){2}}
\put(10.5,0.5){\vector(1,0){1.5}}
%\put(8.5,0.5){\vector(1,1){1}}
\put(9.5,1.5){\vector(-1,-1){1}}
%\put(9.5,1.5){\vector(1,-1){1}}
\put(10.5,0.5){\vector(-1,1){1}}
%\put(9.5,3){\vector(0,-1){1.5}}
\put(9.5,1.5){\vector(0,1){1.5}}
\put(9.65,2.75){$q_\alpha$}
\put(7.25,0.65){$p_\mu$}
\put(11.55,0.65){$p'_\nu$}
\put(9.4,0.2){$ b$}
\put(8.85,1.1){$ s$}
\put(10,1.1){$ q$}
\put(9.4,1.2){$y$}
\put(8.75,0.53){$0$}
\put(10.05,0.53){$x$}
\put(8.20,2.2){$K (K^*)$}
\put(7.35,0.1){${B_s} (B)$}
\put(11,0.1){${B^*} (B_s)$}
\end{picture}
\caption{Perturbative diagrams for the $B_s (B_s)$ off-shell meson (left) and for
$K (K^*)$ off-shell meson (right), for $B_s B^* K $ vertex ($B_s B K^*$ vertex). } \label{fig1}
\end{figure}

The invariant amplitudes receive contributions
from all terms in the OPE. In the case of form factors, the main contribution in the OPE is the 
perturbative term, which is represented in Fig.~\ref{fig1}, for the two cases that we are 
considering, the  $B_s(B_s)$ and $K(K^*)$ meson off-shell for the $B_s B^* K $ ($B_s B K^*$) 
vertex. 

In order to obtain the form factor, we have to choose one of the different structures appearing
in Eqs.(\ref{correBsoff}) and (\ref{correkoff}). As commented above, different structures can 
lead to different results. Therefore, one has to choose the structure less sensitive to higher
dimension condensates, that provide a better stability as a function of the Borel mass, 
and that have a larger pole contribution, when compared with the continuum contribution. This
is considered a ``good'' structure. If there is more than one ``good'' structure, the others 
can also be considered to estimate the uncertainties of the method. 

For $B_s B^* K$ form factor and in the case $B_s$ off-shell meson, we choose the $p'_{\mu} 
p'_{\nu}$ structure, because it satisfies the criteria above. For $K$ off-shell meson, 
we can work with both $p_{\nu}$ and $p'_{\nu}$ structures In this case, we are going to work with the $p'_{\nu}$ structure while the other one will be used for the 
estimate of the uncertainties. The corresponding perturbative spectral densities, which enter 
in Eq.~(\ref{dis}), are :

\begin{equation}
\rho^{(B_s)}(s,u,Q^2)=\frac{3 }{2\pi\sqrt\lambda} \left[ (2m_b -2m_s) E -2 m_b B\right], \, 
\label{rhoBsoff}
\end{equation}
for the $p'_{\mu} p'_{\nu}$ structure of the $B_s$ off-shell case,
and
\begin{equation}
\rho^{(K)}(s,u,Q^2)=-\frac{3}{2\pi\sqrt\lambda} \left[A(p \cdot
p'-2k \cdot p-m_b m_s+m_b^2)+2 \pi(m_b^2-k \cdot
 p')\right],
\label{rhokoff:bsbestk}
\end{equation}
for the $p_{\nu}$ structure in the $K$ off-shell case. In both cases, the quark condensate 
contributions are neglected after the Borel transform.

For $B_s B K^*$ vertex, we use the $p_{\mu}$ structure for $B_s$ off-shell meson and for $K^*$ 
off-shell meson, we have $p_{\mu}$ and $p'_{\mu}$ structures, both giving
excellent sum rules. Again we show the results for one structure and the other is used to 
estimate the uncertainties. The perturbative contribution to the spectral density, when the 
$B_s$ meson is off-shell is :
\begin{equation}
\rho^{(B_s)}(s,u,Q^2)=\frac{3 }{2\pi\sqrt\lambda} \left[A(p \cdot
p'-m_b m_s-2p \cdot k)-p' \cdot k\right] \, \label{rhoBoff}
\end{equation}
 for the $p_{\mu}$ structure. In this case, the quark condensate, $\langle q \bar q \rangle$, contribution to 
the same structure is:
\begin{equation} \Pi^{\langle q \bar q \rangle}= \frac{ m_s \langle q \bar q \rangle} { (p^2 -m_b^2) ({p'}^2-m_s^2)}. 
\end{equation}
For $K^*$ off-shell case, the spectral density, for  both structures, is given by:
$$ \rho^{(K^*)}(s,u,Q^2)=-\frac{3}{2\pi\sqrt\lambda}
\{p_{\mu}\left[A(p \cdot p'-m_b m_s-m_b^2)+m_b^2-k \cdot
 p'-m_b m_s)\right]$$
\begin{equation} +p'_{\mu}\left[B(p \cdot p'+m_b m_s-m_b^2)-k \cdot
 p+m_b^2\right]\}.
\label{rhokoff:bsbkest}
\end{equation}

In Eqs.~(\ref{rhoBsoff}) to (\ref{rhokoff:bsbkest}), we have defined
$\lambda = \lambda(s,u,t) = s^2+t^2+u^2-2st-2su-2tu$, $s=p^2$,
$u=p'^2$, $t=-Q^2$ and $A, ~B$ and $E$  are functions of $(s,u,t)$,
given by:
\begin{eqnarray}
A=2 \pi \left[
\frac{\overline{k_0}}{\sqrt{s}}-\frac{\overline{|\overrightarrow{k}|}}{|\overrightarrow{p'}|}
\overline{cos
\theta}\frac{p'_0}{\sqrt{s}} \right]  \label{D};
\;\;\;\;\;\;\;\;\;\;\; B= 2\pi\frac{\overline{|\vec k|}}{
\overline{|\vec p'|}} \overline{\cos\theta} \,\, \label{N} ;
\end{eqnarray}
\begin{eqnarray}
E=-\frac{\pi\overline{|\vec k|}^2 }{\overline{|\vec p'|}^2}
\left(3 \overline{\cos\theta} -1 \right),
\end{eqnarray}
where
\begin{eqnarray}
\overline{|\vec k|}^2&=& \overline{k_0}^2-m_i^2   \label{vk}; \;\;\;\;\;\;
\overline{\cos\theta}=-\frac{2p'_0\overline{k_0}-u-m_i^2 -\eta m_b^2}
{2 \overline{|\vec p'|} \overline{|\vec k|}}   \label{ctheta};\nonumber \\
p'_0&=&\frac{s+u-t}{2\sqrt{s}}    \label{pl0}; \;\;\;\;\;\;\;
\overline{|\vec p'|}^2=\frac{\lambda}{4s}     \label{vpl};\;\;\;\;\;\;\
\overline{k_0}= \frac{s+ m^2_i-\epsilon \; m_b^2}{2\sqrt{s}}   \label{k0b} ;
\end{eqnarray}

Finally, the OPE side, is calculated using  Eq.~({\ref{dis}}) with the limits in the 
integration given by: $s_{min}=(m_b +m_s)^2 $ and $u_{min}= t -m_b^2$ for $B_s$ off-shell and
$s_{min}=m_b^2-m_s^2$ and $u_{min}= t + m_b^2-m_s^2$ for $K$ off-shell, for $B_s B^* K$ vertex.
And $s_{min}=(m_b )^2 $ and $u_{min}= t + m_b^2$ for $B_s$ and  $K^*$ off-shell for the  
$B_s B K^*$ vertex.

\subsection{The phenomenological side}

 The three-point functions from Eqs.~(\ref{correBsoff}) and
(\ref{correkoff}), when written in terms of hadron masses, decay
constants and form factors, give the phenomenological side of the
sum rule. 

\subsubsection{ For the $B_sB^*K$ vertex:} 
The meson decay constants $f_{K}$, $f_{B_s}$ and $f_{B^*}$  are
defined by the following  matrix elements:
 \beq \langle
0|j_{\nu}^{K}|{K(p)}\rangle= i  f_{K} p_{\nu}, \label{fK:bsbestk} \eeq
\beq \langle 0|j_{\mu}^{B^{*}}|{B^{*}(p')}\rangle= m_{B^{*}}
f_{B^{*}} \epsilon_{\mu}^*(p') \, , \label{fBss:bsbestk} \eeq
and \beq \langle 0|j_{5}^{B_s}|{B_s(p)}\rangle=
\frac{m_{B_s}^2}{m_b+m_s} f_{B_s}  \, , \label{fB:bsbestk} \eeq
and the vertex function is defined by: 
\beq
\langle  K(q) |B^{*}(-p'){B_s(p)} \rangle = -g^{(K)}_{B_s B^*
 K} \epsilon^{{\mu}*}(p') (2 p- p')_{\mu} \, ,
\label{vertex:bsbestk} \eeq
which is extracted from the effective Lagrangian  \cite{lagrangiana}:
\beq \mathcal{L}_{B_s B^* K}= ig_{B_s B^* K} \Big [  B^{*\mu} (
\bar{B_s}
\partial_{\mu}K - \partial_{\mu}\bar{B_s} K ) -  \bar{B^{*\mu}} (
B_s
\partial_{\mu}\bar{K} - \partial_{\mu} B_s \bar{K}  )    \Big ].\;,
\nonumber \label{lagr:bsbestk}
\eeq
Saturating the correlation function with $B_s,~B^*$ and $K$ intermediate states we
arrive at
 \beqa &&\Pi_{\mu \nu
}^{(B_s)}=\frac{- f_K f_{B_s} f_B
m_{B^*}m^2_{B_s}g^{(B_s)}_{B_sB^*K}(q^2)}{(m_s+m_b)(p^{\prime 2}-m_K^2)(q2-m^2_{B_s})(p^2-m^2_{B^*})}\nonumber\\
 &&\times \left[p_{\mu}{p^{\prime}}_{\nu} \left(1-\frac{(m_{K}^2-q^2)}{m_{B^*}^2}\right) -2{p^{\prime}}_{\mu}{p^{ \prime}}_{\nu}\right] + {\textit{``continuum''}},
\label{phenBoff} \eeqa
for an off-shell $B_s$. Using the matrix element of $K$ meson
equal to \beq \langle 0|j_{5}^{K}|{K(q)}\rangle=
\frac{m_{K}^2}{m_s} f_{K}  \, , \label{fK2} \eeq
we arrive at an expression for an off-shell $K$:
\beqa &\Pi^{(K)}_{\mu }&=\frac{-f_{B^*}  f_K
f_{B_s} m_{B^*} m_{B_s}^2 m_K^2  g^{(K)}_{B_s B^*K}(q^2)}{(m_b+m_s) m_s(p^2-m_B^2) ({p^{\prime}}^2+m_{B_s^*}^2)
(q^2-m^2_K)}\nonumber \\
&&\left[-2p_{\mu}+{p ^{\prime}}_{\mu} \left(1+\frac{(m_{B_s}^2-q^2)}{m_{B^*}^2}\right)\right]
  +{\textit{``continuum''}}. \label{phenKoff} \eeqa

	\subsubsection{For the $B_s B K^*$ vertex:}
	
	In this case, the effective Lagrangian is \cite{lagrangiana}:

\beq \mathcal{L}_{B_s BK^*}= ig_{B_s BK^*} \Big [  K^{*\mu} ( B
\partial_{\mu}\bar{B_s} - \bar{B_s}\partial_{\mu}B ) +  \bar{K^{*\mu}} (
B_s
\partial_{\mu}\bar{B} - \bar{B}\partial_{\mu} B_s) \Big ]\;,
\nonumber \label{lagr:bsbkest} \eeq 
from where we can extract the vertex element, which is given by:
 \beq
\langle  K^*(q) |B_s(-p')B(p) \rangle = ig^{(K^*)}_{B_s BK^*}
\epsilon^{*\mu}(q) (p+ p')_{\mu} \, , \label{vertex:bsbkest} \eeq 
and the matrix elements which introduce the meson decay constants $f_{K^*}$, $f_{B_s}$ and $f_{B}$ are:

 \beq \langle
0|j_{\mu}^{K^*}|{K^*(p)}\rangle= f_{K^*}
\epsilon_{\mu}^*(q)m_{K^*}, \label{fK:bsbkest} \eeq

\beq \langle {B(p)|j_5^B|0}\rangle= f_{B} \frac{m^2_B}{m_b^2}  \,
, \label{fBss:bsbkest} \eeq

and \beq \langle 0|j_{5}^{B_s}|{B_s(p')}\rangle=
 f_{B_s}\frac{m_{B_s}^2}{m_b+m_s}  \, , \label{fB:bsbkest} \eeq

After some algebra we arrive at the following expression:

\beqa &\Pi^{(B_s)}_{\mu}&=-i\frac{f_{K^*} f_B f_{B_s}
m_{K^*} m_{B_s}^2 m_B^2}{(m_b^2 +m_s m_b)(p^2-m_{B}^2) ({p
\prime}^2-m_{K^*}^2)
(q^2-m^2_{B_s})}\nonumber \\
&&\times g_{B_s BK^*}^{(B_s)}(q^2)\left[-2p_{\mu}
%\right.\nonumber \\ &&\left.
+{p \prime}_{\mu} \left(1-\frac{m_B^2 -m_{B_s}^2}{m_{K^*}^2}\right)\right]+
{\textit{``continuum''}},
\label{phenBsoff} \eeqa
when $B_s$ is off-shell. 

For $K^*$ off-shell we arrive at:

\beqa &\Pi^{(K^*)}_{\mu}(p,p \prime,q)&=-\frac{f_B f_{K^*}
f_{B_s}m_{K^*} m_{B_s}^2 m_B^2}{(m_s m_b+m_b^2)(p^2-m_B^2) ({p
\prime}^2-m_{B_s}^2)
(q^2-m^2_{K^*})}\nonumber \\
&&\times g_{B_s BK^*}^{(K^*)}(q^2)\left[p_{\mu}\left(1-\frac{(m_B^2-m_{B_s}^2)}{m_{K^*}^2}\right)\right.\nonumber \\
 &&\left.+{p \prime}_{\mu} \left(1-\frac{(m_{B_s}^2-m_B^2)}{m_{K^*}^2}\right)\right]+{\textit{``continuum''}} \label{phenKsoff} \eeqa

\section{The sum rule}

The sum rule is obtained after performing a double Borel transform ($\cal BB$), $P^2 = 
- p^2 \to M^2$ and $P'^2 = -p'^2 \to M'^2$, to both the phenomenological and OPE sides: 

\begin{equation}
{\cal BB}\left[\Gamma_\mu^{OPE(I)}\right](M,M')={\cal BB}\left[\Gamma_\mu^{phen(I)}\right](M,
M')\,, \label{qhb}
\end{equation}
where $M$ and $M'$ are the Borel masses and $I$ is the off-shell meson.
 
In order to eliminate the continuum contribution in the phenomenological side, instead of doing the 
integrals in Eq.~(\ref{dis}) up to $\infty$, we do the integrals up to the continuum threshold 
parameters $s_0$ and $u_0$. The threshold parameters are defined as $s_0 = (m_i + \Delta_i)^2$ 
and  $u_0 = (m_o + \Delta_o)^2$, where $\Delta_i$ and $\Delta_o $ are usually taken as $0.5$ 
MeV, and $m_i$ and $m_o$ are the masses of the incoming and outgoing mesons respectively.

In Eqs.~(\ref{phenBoff}), (\ref{phenKoff}), (\ref{phenBsoff}), (\ref{phenKsoff}), 
 $g^{(I)}_{B_sB^* K}(Q^2)$  and $g^{(I)}_{B_sB K^*}(Q^2)$ are the form factors when the $I$ 
meson is off-shell. As in our previous works, we define the coupling constant as the value of 
the form factor, $g^{(I)}(Q^2)$, at $Q^2 = -m^2_I$, where $m_I$ is the mass of the off-shell 
meson.

\section{Results and discussion}

 Table \ref{param} shows the value of the hadronic parameters used in the present
calculation. We have used the experimental value for $f_{K}$ of
Ref.~\cite{fkvalue}, for $f_{B^*}$  and $f_{B_s}$ from Ref.~\cite{fbsvalue} and 
for $f_{K^*}$ of  Ref.~\cite{fke}.
\begin{table}[ht]
\begin{center}
\caption{Hadronic parameters used.}
\begin{tabular}{cccccc}\hline
   & $K$ & $K^*$& $B_s$ & $B^*$ &  $B$\\\hline
  $m$ (GeV)  & $0.49$ & $0.89 \pm 5$ & $5.40$ & $5.20$ & $5.28$\\
  $f$ (MeV)&$160^{\pm 1.4}_{\pm 44}$ & $220\pm 5 $& $208^{\pm 10}_{\pm 39}$ &$250 ^{\pm10}_{\pm29}$ & $191 \pm 0.87 $\\\hline
\end{tabular}
\label{param}
\end{center} 
\end{table}

We neglect the light quark mass ($m_q = 0.0\MeV$). The strange and bottom quark masses  were taken from the 
Particle Data Group (PDG) and have the values $m_s = 104 + 26 - 34\MeV$ and $m_b = 4.20+0.17-0.07\GeV$ respectively.  In the next subsections, we show the two form factors  used to extract the coupling constants  of each vertex.

%------------------ For B_s^*BK meson off-shell

\subsection{ $ B_s B^*  K$ vertex }

%\subsubsection{ $B_s$ off-shell form factor}
\begin{figure}[ht!]
\centering
\includegraphics[width=0.49\linewidth]{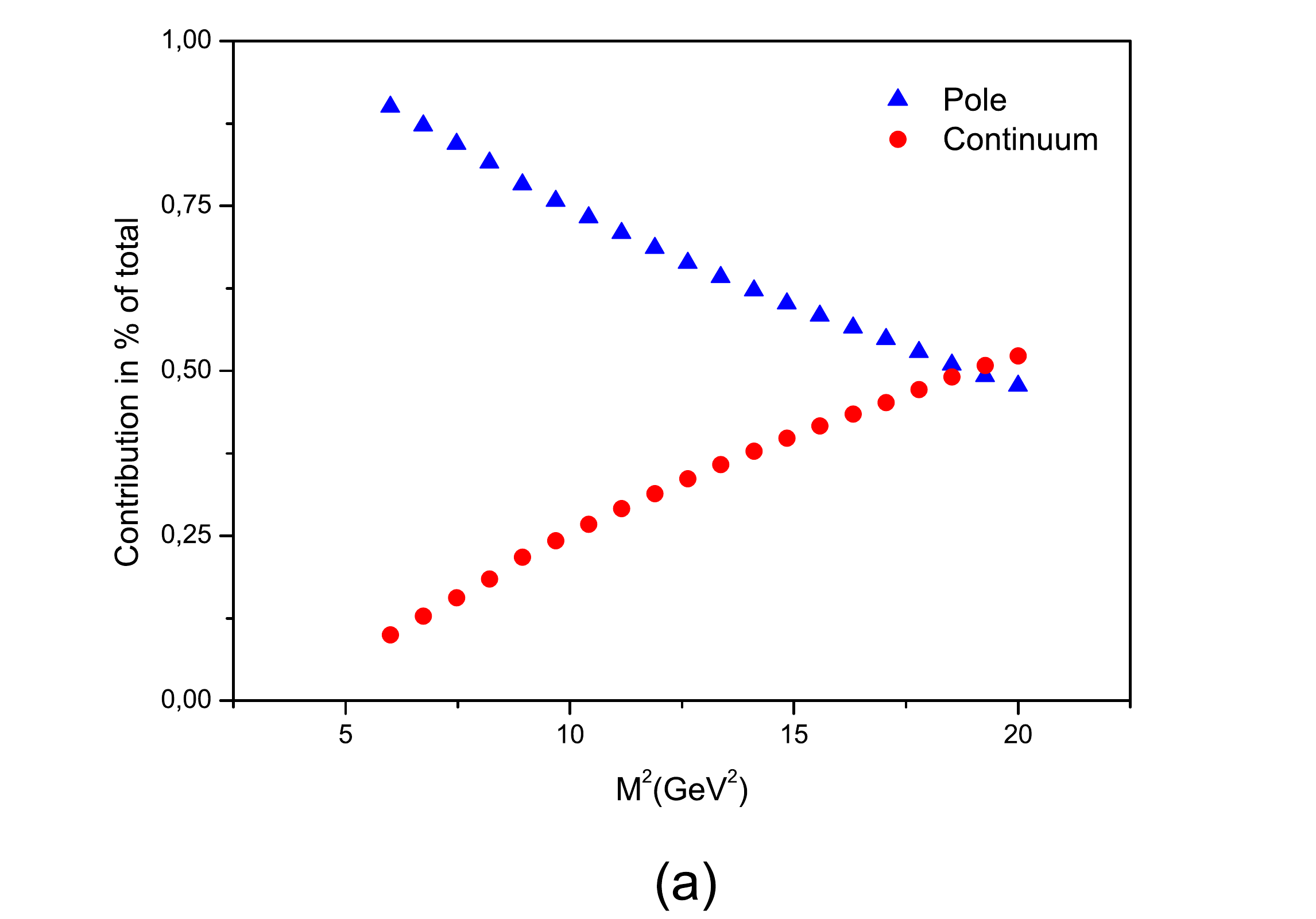}
\includegraphics[width=0.49\linewidth]{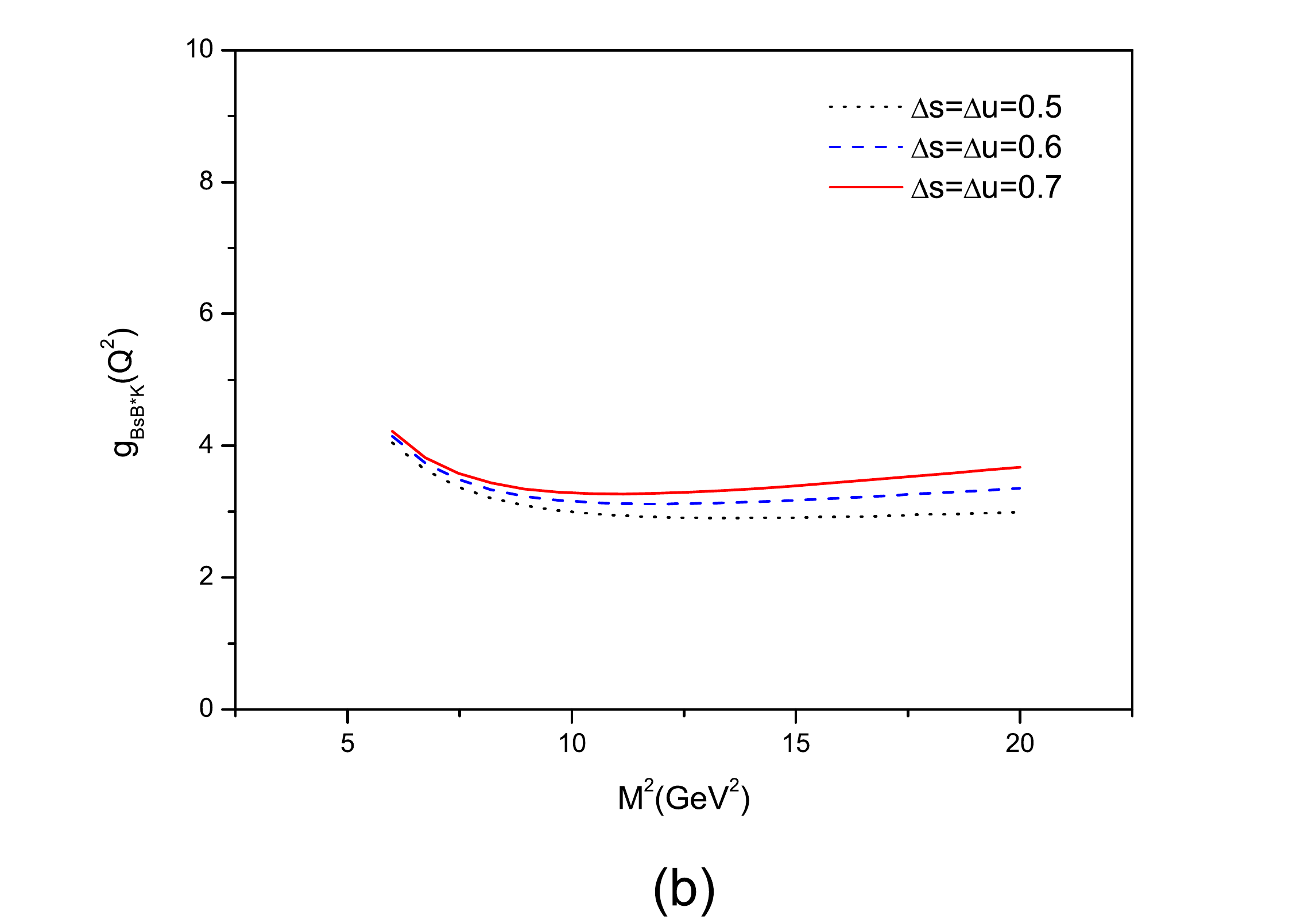}
\caption{(a) The pole-continuum
contributions for the $ B_s B^* K$ sum rule for the $B_s$ off-shell and (b) 
$g^{(B_s)}_{B_s B^* K}(Q^2=1 \GeV^2)$ stability for different values of the continuum 
thresholds.} 
\label{estabilidadeBoff}
\end{figure}

In the case of a $B_s$ off-shell meson, we work with the  $p'_{\nu}p'_{\mu}$ structure. 
In Fig.~\ref{estabilidadeBoff}(a) we show the contribution of the pole versus the continuum 
contribution for the sum rule and in Fig.~\ref{estabilidadeBoff}(b) the stability of the form factor as a function o 
Borel mass, for $Q^2=1$ GeV$^2$ and three different values for the continuum thresholds. 
We use the usual relation between the Borel masses $M'$ and $M$ \cite{Bracco:2011pg}: 
$ M'^{2}= \frac{m^2_K}{m_{B^*_s} -m^2_b} M^2 $. 
 
From Fig.~\ref{estabilidadeBoff}(b) we clearly see a window of stability for  $M^2 \geq 10 \; 
GeV^2$, and from Fig.~\ref{estabilidadeBoff}(a), we can see that the pole contribution is bigger
than the continuum contribution for $M^2\leq19 \GeV^2$. Therefore, there is a Borel window
where the sum rule can be used to extract the form factor.

%\subsubsection{ $K$ off-shell form factor} 
%------k  off P ---------------
\begin{figure}[ht]
\centering
\includegraphics[width=0.49\linewidth]{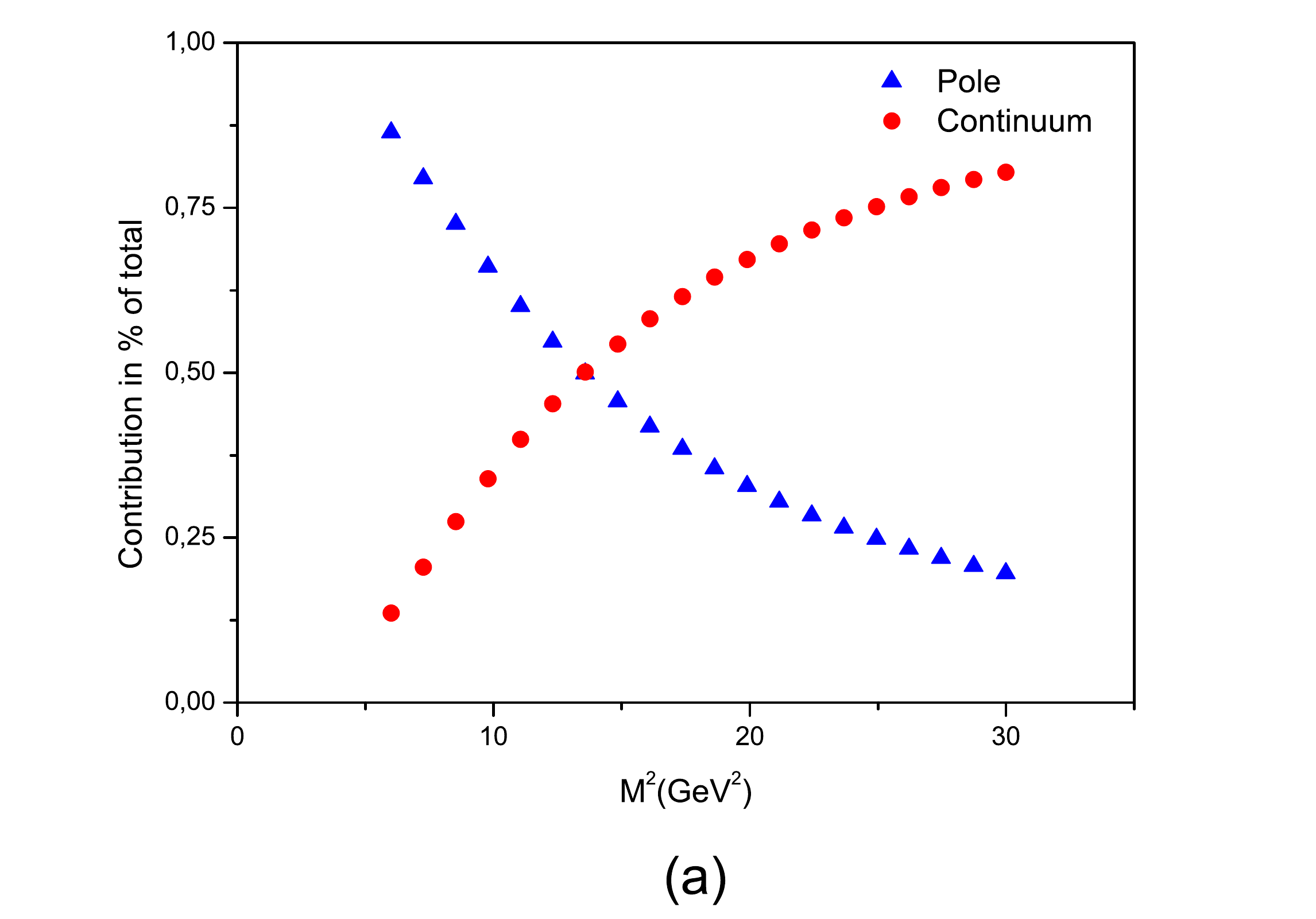}
\includegraphics[width=0.49\linewidth]{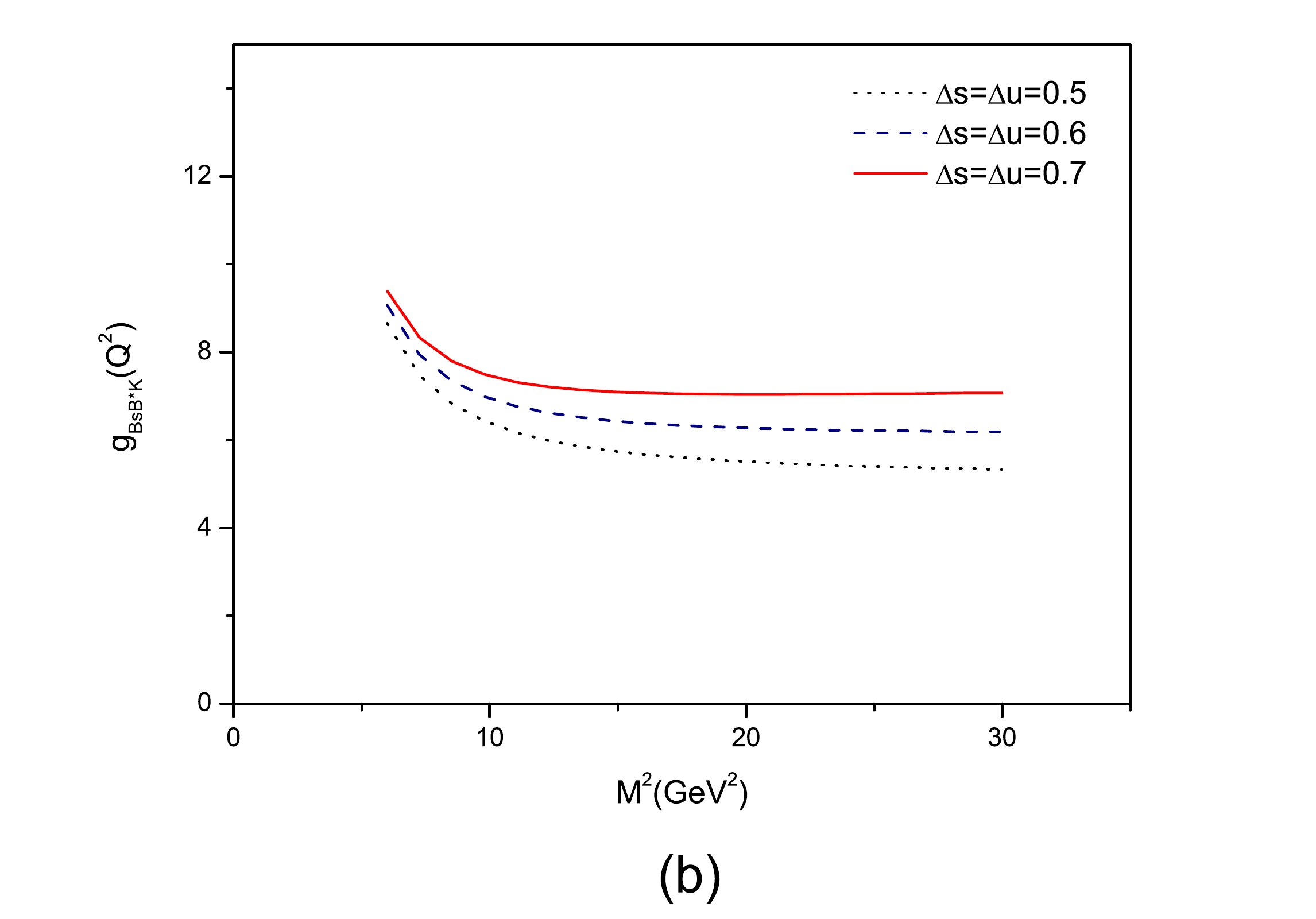} 
\caption{a) The pole and continuum contributions for ${B_s B^* K}$ sum rule with the $K$ 
off-shell and b) the $g^{(K)}_{B_s B^* K}(Q^2 = 2 \; GeV^2, M^2)$ stability  for different 
values of the continuum thresholds}
\label{estabilidadeKoffp}
%\end{center}
\end{figure}

In the case of a $K$ meson off-shell, we have two structures in
Eq.~(\ref{phenKoff}) that can be used:  $p'_{\mu}$ and $ p_{\mu}$.
Both structures give good sum rule results, that means a good
pole-continuum contribution and good stability. 
We show in Fig.~\ref{estabilidadeKoffp}(a) and (b), repectively, the pole-continuum 
contribution  and the stability only for the $p_{\nu}$ structure.  For the $p'_{\mu}$, we 
obtain a very similar result and we use it to evaluate the uncertainties.

From Fig.~\ref{estabilidadeKoffp}, we find a Borel window $10\GeV^2\leq M^2\leq 14
\GeV^2$ where the sum rule can be used to extract he form factor.

In Fig.~\ref{formfactorBsBeK} we show the QCDSR results for these two form factors, 
represented by squares and triangles for the cases of the $B_s$ and $K$ off-shell
respectively.

%And for the Structure $P'$.
%%%------k  off P' ---------------
%\begin{figure}[ht]
%%\begin{center}
%{\epsfig{figure=polo-conti-K-pl.eps,height=55mm}}
%{\epsfig{figure=estabilidade-K-pl.eps,height=55mm}} \caption{a)
%$g^{(K)}_{B^*_s BK}(Q^2 = 2 \; GeV^2)$ stability as a function of
%the Borel mass for different threshold values, structure $
%p^\prime_{\nu}$ and b) pole and continuum contributions. }
%\label{estabilidadeKoffpl}
%\end{center}
%\end{figure}
%========================================================================
\subsection{ $ B_s B K^*$ vertex }

%\subsubsection{ $B_s$ off-shell form factor}

\begin{figure}[t]
\centering
\includegraphics[width=0.49\linewidth]{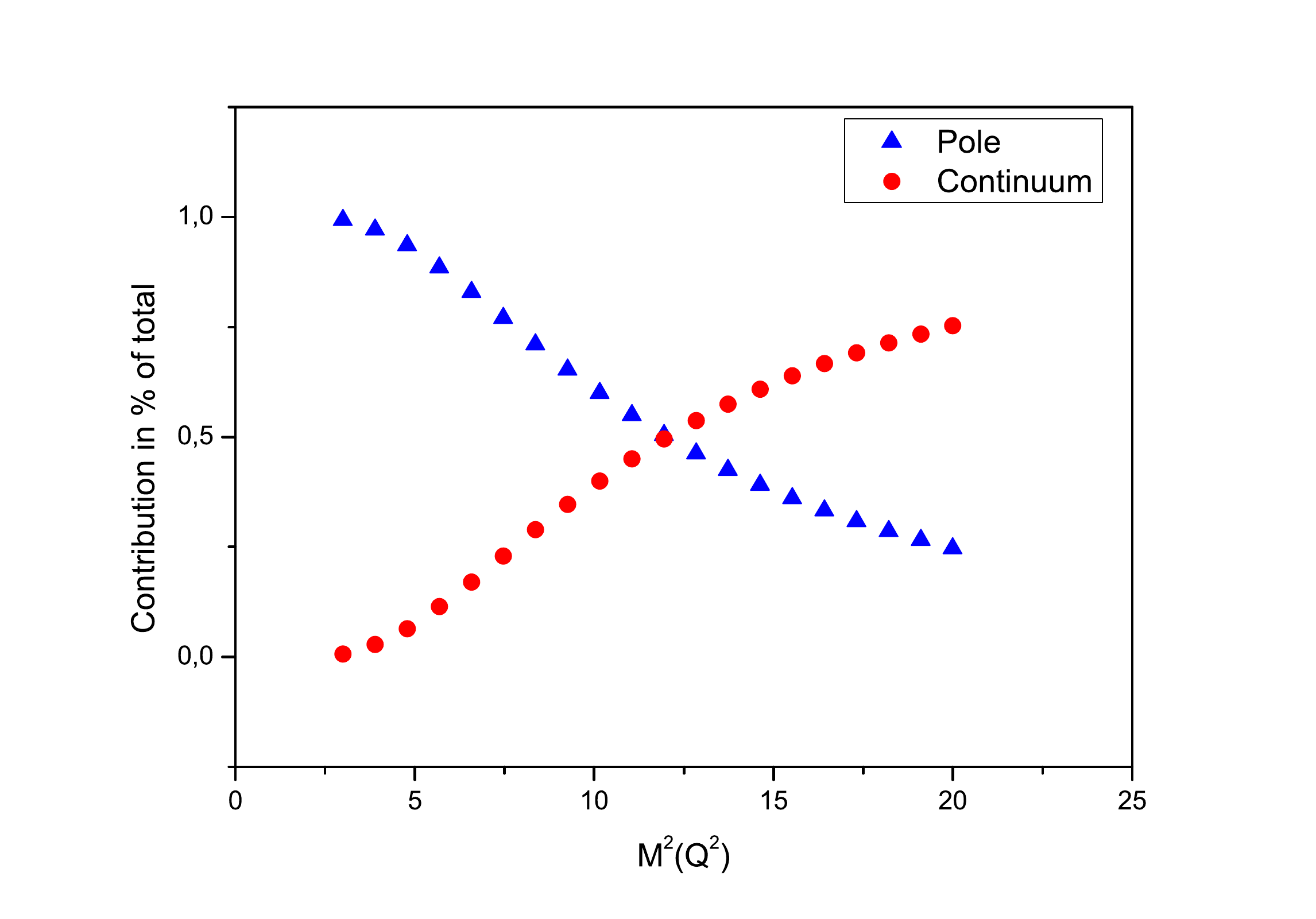}
\includegraphics[width=0.49\linewidth]{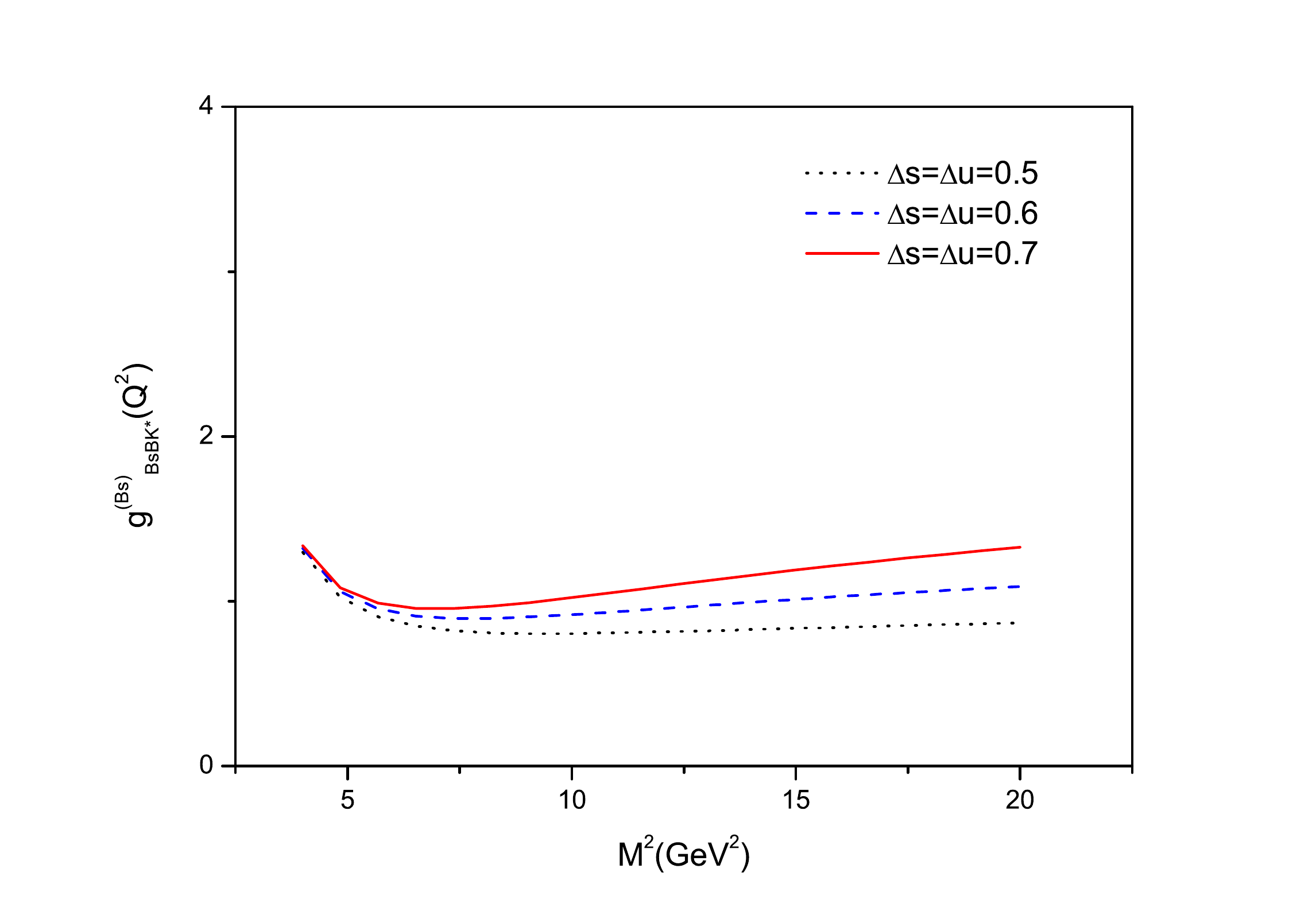}
\caption{a) The pole-continuum contributions for $ B_s BK^*$ and in b) 
$g^{(B_s)}_{B_s BK^*}(Q^2=1 \GeV^2, M^2)$ stability for different thresholds.} 
\label{estabilidadeBoff-v2}
%\end{center}
\end{figure}

For the case of the $B_s$ meson off-shell, we use the structure $p_{\mu}$. 
In Fig.~\ref{estabilidadeBoff-v2}(a) and (b) we show  the pole-continuum contribution 
contribution and the Borel mass stability respectively. Once more, we see that we get a Borel 
window $10 \GeV^2\leq M^2\leq 12.5\GeV^2$ where these both conditions are satisfyed and
where we can use the sum rule to extract the form factor.
Also, for the $B_s$ meson off-shell case, we have other structure,  $p'_{\nu}$,
to work with. In this case, the result includes the light quarks condensate which do not
modified substantially the sum rule. The $p'_{\nu}$ structure will be used to evaluate the uncertainties.

%\subsubsection{ $K^*$ off-shell form factor}

\begin{figure}[ht]
\centering
\includegraphics[width=0.49\linewidth]{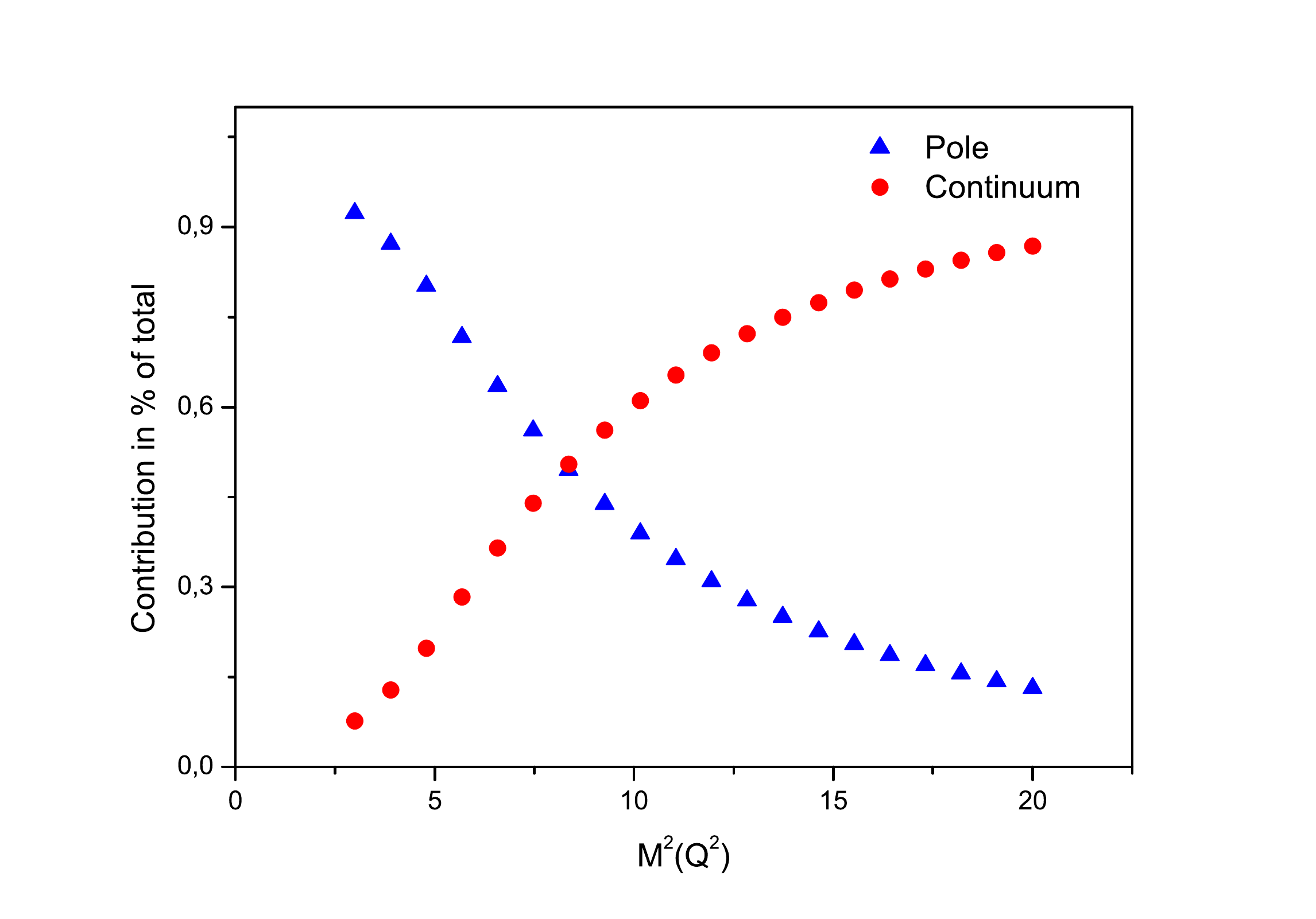}
\includegraphics[width=0.49\linewidth]{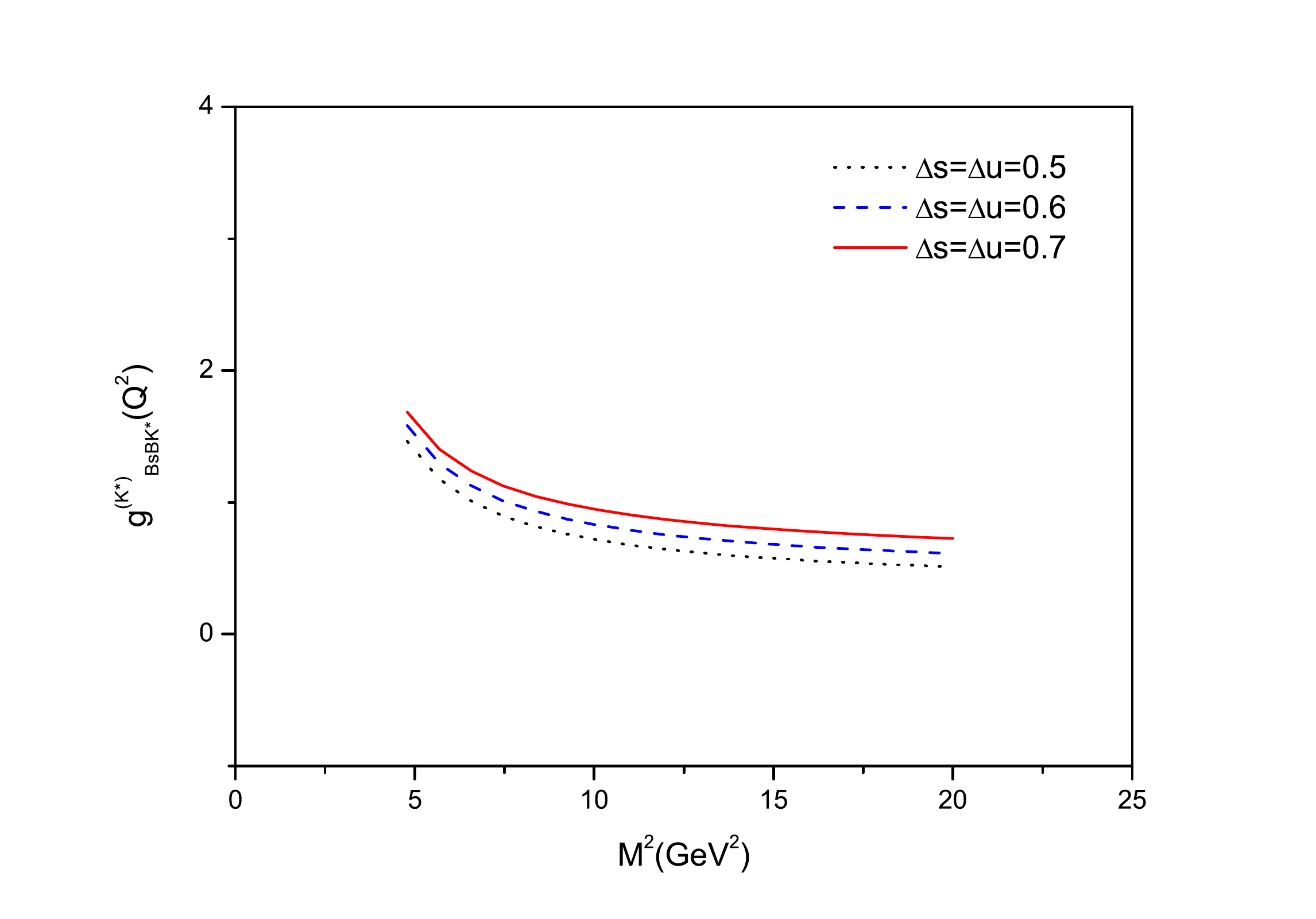}
\caption{a) pole and continuum contributions for $B_s BK^*$ sum rule for $K^*$ off-shell
and b) the $g^{(K^*)}_{B_s BK^*}(Q^2 = 2 \; GeV^2, M^2)$ stability for different values of the
threshold parameters}
\label{estabilidadeKoffp-v2}
%\end{center}
\end{figure}

In the case of the $K^*$ off-shell meson, we show the figures of stability and pole-continuum 
contribution for $p'_{\mu}$ structure, in Fig.\ref{estabilidadeKoffp-v2}. In the structure
$ p_{\mu}$, we get similar results, that means a good pole-continuum contribution and good 
stability. 

The sum rule result for both form factors are shown in Fig.\ref{formfactorBsBKe} through
triangles and squares for $B_s$ and $K^*$ off-shell respectively.

\section{Coupling constant}

The coupling constant in the vertex, is obtained when the form factor is 
extrapolated to $Q^2= -m^2_{I}$, where $m_{I}$ is the mass of the
off-shell meson.  In order to minimize the uncertainties, we work with two form factors in each
vertex, one with the heavier meson in the vertex off-shell and the other with the lighter 
meson off-shell. Both form factors should give the same value for the coupling constant. 

In Table \ref{couplings}, we show, for each vertex, the function that fits the QCD sum rule 
results for the form factors, which is then extrapolated to  $Q^2= -m^2_{I}$, to determine 
the coupling constant value. From Table  \ref{couplings} we see that
the two form factors in the same vertex, give similar results for the coupling constant.

\begin{table}[ht]
\begin{center}
\caption{Form factors and coupling constants obtained by extrapolation}
\begin{tabular}{lc}\hline
    Form Factor &  Coupling Constant \\\hline
$g_{B_s B^* K}^{(B_s)}(Q^2)=3.18\exp^{-Q^2/32.98}$ & $g_{B_s B^* K}= 7.55  $ \\
$g_{B_s B^* K}^{(K)}(Q^2)= 8.25 \exp^{- (Q^2/6.68)^2}$ &$g_{B_s B^* K}= 8.00  $      \\    
$	g_{B_s B K^*}^{(B_s)}(Q^2)=\frac{80.88}{58.86+Q^2}$  & $g_{B_s B K^*}= 2.72$ \\
$g_{B_s BK^* }^{(K^*)}(Q^2)= 2.49 \exp^{- Q^2/2.63}$&  $g_{B_s BK^*}= 3.37 $   \\
\hline
	\end{tabular}
\label{couplings}
\end{center}
\end{table}

% figura dos fatores de forma para BsBeK

\begin{figure}[h]
\centering
\includegraphics[height=80mm]{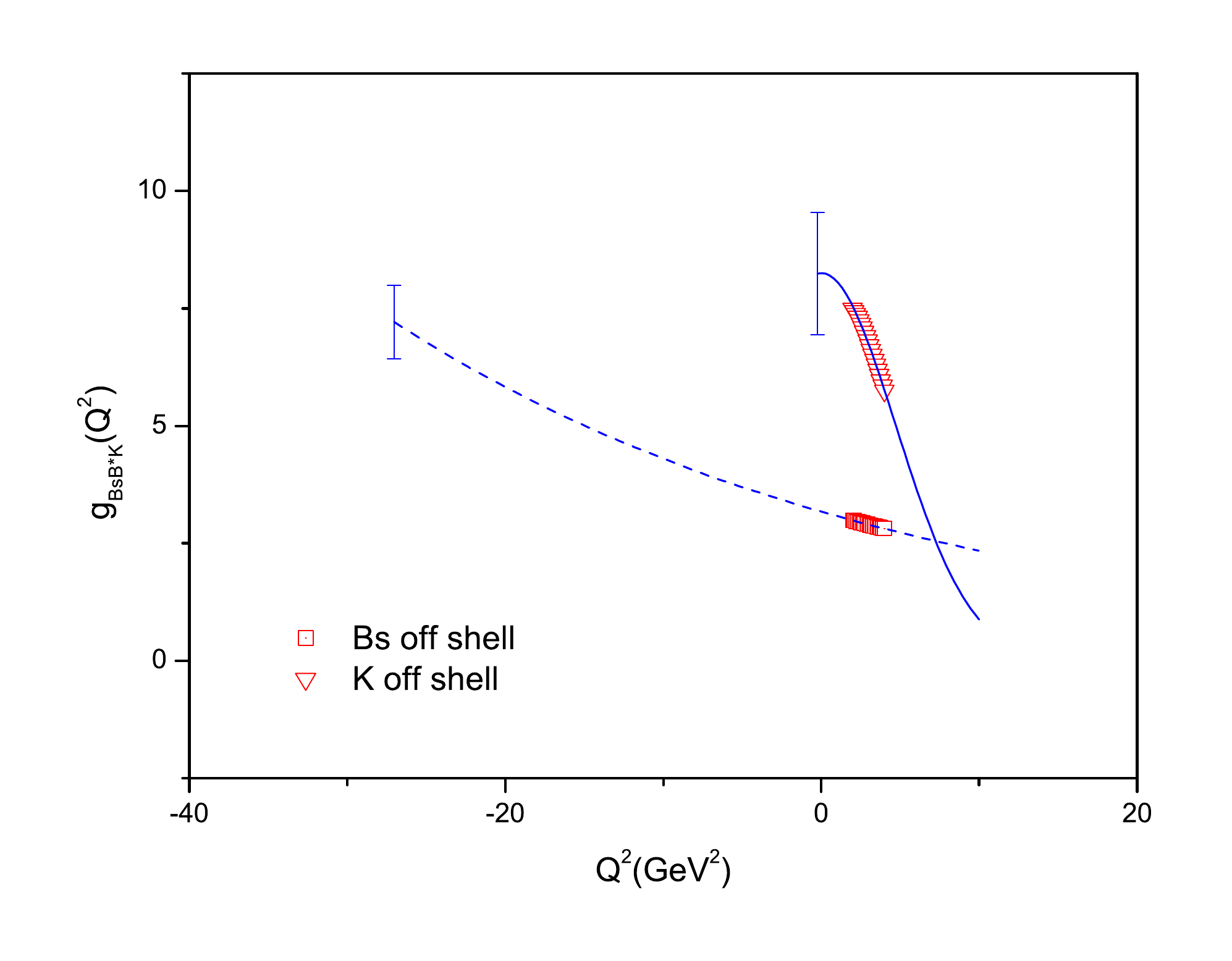}
\caption{The $g_{B_s B^* K}(Q^2)$ form factor, extrapolation function for $K$ and $B_s$
meson off-shell.} 
\label{formfactorBsBeK}
%\end{center}
\end{figure}

\begin{figure}[h,t]
\centering
\includegraphics[height=80mm]{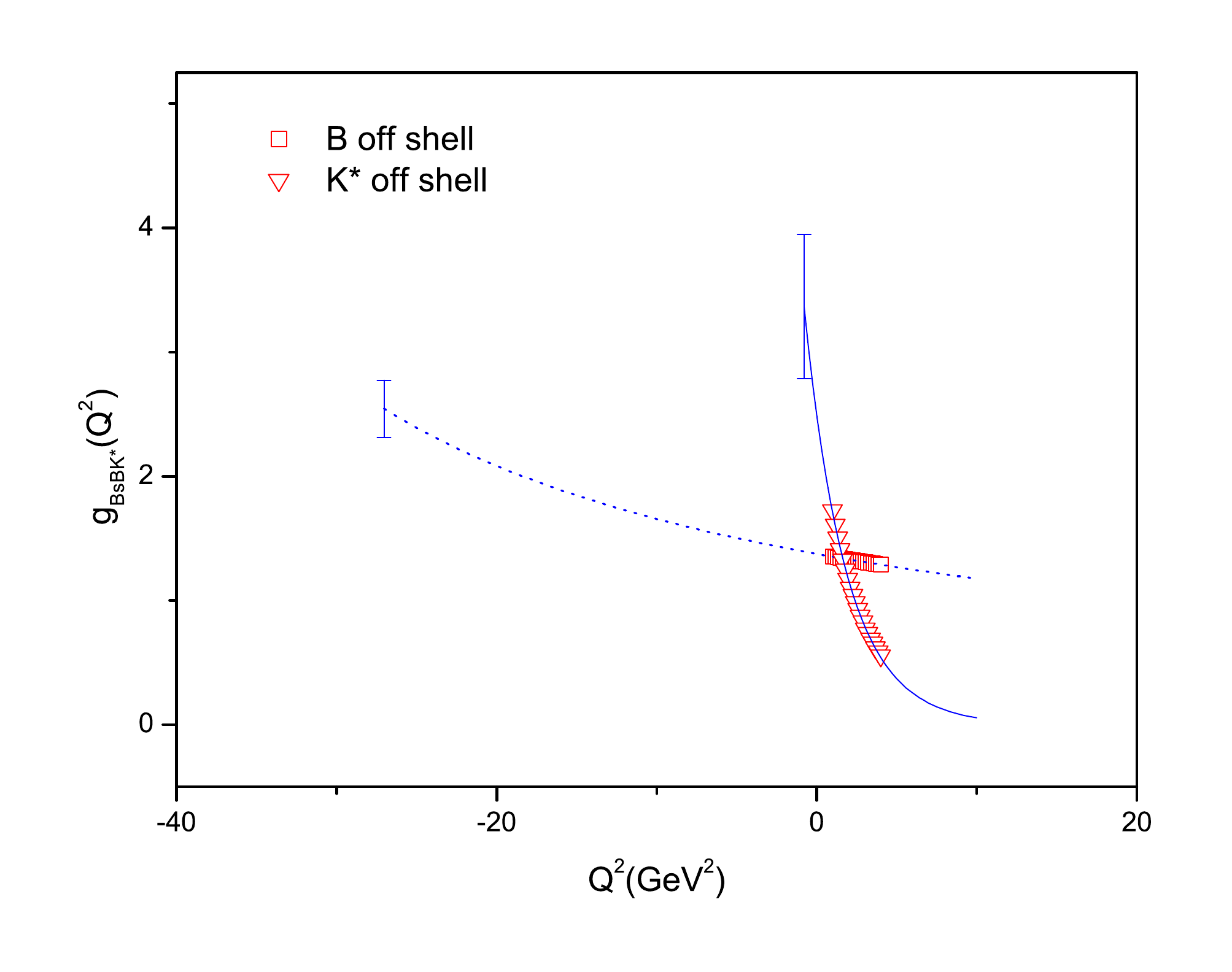}
\caption{The $g_{B_s B K^*}(Q^2)$ form factor, extrapolation function for $K^*$ and $B_s$
meson off-shell.} \label{formfactorBsBKe}
%\end{center}
\end{figure}

In Figs.~\ref{formfactorBsBeK} and \ref{formfactorBsBKe}, we show the parametrizations
given in Table \ref{couplings} for the QCDSR results for the form factors 
$g_{B_s B^* K}(Q^2)$ and $g_{B_s B K^*}(Q^2)$ respectively. The squares and triangles
give the QCDSR  results for the heavier and lighter off-shell mesons in the vertices 
respectively.
We also show in these figures the values obtained for the coupling constant.

%For $B_s B K^*$ 

\section{Uncertainties of the QCDSR method}

Looking at Fig. \ref{formfactorBsBeK} and \ref{formfactorBsBKe}, we can see a  error bar at the 
endpoint of the curves. To evaluate these error bars, we compute the sum rule taking into 
account the errors in the masses, decay constants, continuum threshold parameters  and also we 
study a variation on the Borel mass in the window of $Q^2$. 

Our procedure is that in each computation, all the parameters are kept fixed and only one 
changes within its intrinsic error.  In Table \ref{desvios}, we show the intrinsic uncertainty
for each of the sum rule parameters.

%---------------------------------------- Tab.(2)-----------------

\begin{table}[h!]
\begin{center}
\caption{Percentage deviation of the coupling constant related to each parameter for both vertices with the meson $I$ off-shell. }
\begin{tabular}{lcccc}\hline
    &    \multicolumn{4}{c} {\textbf{Deviation {\%}}} \\
		&    \multicolumn{2}{c} {\textbf{$g_{B_s B^* K}^{(I)}$}}  & \multicolumn{2}{c} {\textbf{$g_{B_s B K^*}^{(I)}$}} \\\hline
\textbf{Parameters} &\textbf{ $I = B_s$}& \textbf{$I = K$}& \textbf{$ I = B_s$} & \textbf{$I = K^*$} \\ \hline 
  $f_{K^*}=220 \pm 5$(MeV)                 &     --      &  --   &   $1.95$    &$1.94$ \\
	$f_{B}=191 \pm 8.77$(MeV)                &     --      &  --   &    $3.91$      &$3.88$  \\
  $f_K =159.8 \pm 1.4 \pm 0.44 $(MeV)      & $1.04$    & 1.04 &    --        & --\\
  $f_{B^*}=208 \pm 10 \pm 29 $ (MeV)       & $16.29$ & $16.30 $ & -- &   --    \\
  $f_{B_s}=250 \pm 10 \pm 35 $ (MeV)       &  $12.06$ & $12.07$ & $3.61$      & $3.51$  \\
  $m_b=4.20+0.17-0.07$ (GeV)               & $15.81$ & $23.39$&$9.83$       & $28.42$       \\
  $m_s=104 + 26 - 34$ (MeV)                & $1.23$   &$24.71$ & $7.21$       & $19.78$ \\
	 $M^2 \pm 10\% $ (GeV)                    & $4.29$ & $ 1.85$  &$6.45$       & $ 18.57$          \\
  $\Delta s \pm 0.1$ e $\Delta u \pm 0.1 $(GeV)  & $13.51$ &$ 4.90$& $9.32$ &$12.35$\\
	\hline
\end{tabular}
\label{desvios}
\end{center}
\end{table}

We also consider the other good sum rules, for each off-shell meson in each vertex and 
we also put the third meson off-shell to better estimate the errors.

\subsection{Other mesons of shell and other structures}

For the ${B_s B^* K}$ vertex, we calculated the $B^*$ off-shell form factor. 
Since  the $B^*$ meson has a small
mass difference when compared with the $B_s$ meson mass, we expect
to obtain a similar result to the one obtained for $B_s$ off-shell.
Computing the sum rule, we observe that there are two structures:
 the $\pli_{\mu}\pli_{\nu}$ and $p_{\mu} \pli_{\nu}$.
The $p_{\mu}\pli_{\nu}$ structure does not give a good stability
in the Borel mass, therefore it is not considered. The other
structure, $\pli_{\mu}\pli_{\nu}$, gives a good stability and the pole
is larger than the continuum contribution.

%Finally, the uncertainties for each coupling constant, which is
%obtained with $B_s$, $K$ and $B^*$ off-shell are: \beq
% g^{(B_s)}_{B_s^* BK}=7.55 \pm 0.78
%\eeq for $B_s$ off-shell meson,
%\beq
% g^{(K-p)}_{B_s B^* K}=8.0 \pm 1.3
%\eeq
%\beq
% g^{(K-pl)}_{B_s B^* K}=9.45 \pm 1.51
%\eeq
%for $K$ off-shell,
%where were used both structures that give good SR,
%and \beq g^{(B^*)}_{B_s^* BK}=8.64\pm 1.31 \eeq for $B^*$
%off-shell.
To extrapolate the QCDSR results, we use the fit
\begin{equation}
g_{B_s B^* K}^{(B^*)}(Q^2)= 1.21\exp^{-Q^2/13.65} \;
\label{expBxoff}
\end{equation}
for the form factor, and the resulting value for the coupling constant is:
$g_{B_s B^* K}= 8.64 \label{couplingbssoff}$.  This result is very similar to the one
showed in Table \ref{couplings}, as expected.

Also for this vertex, in the case of the $K$ meson off-shell, we have other structure, 
$p'_{\mu}$, which gives a good sum rule, very similar to the $p_{\mu}$ used
before.  In this case the form factor is extrapolated by 
\begin{equation}
g_{B_s B^* K}^{(K)}(Q^2)= 9.69\exp^{-(Q^2/7.43)^2} \;,
\label{koffBsBeKpmu}
\end{equation}
and the resulting value for the coupling constant is:
$g_{B_s B^* K}= 9.45 \label{koffBsBeKcoupling}$, in a very good agreement with the results in 
Table \ref{couplings}.

For $B_s B K^*$ , we perform the sum rule for the $B$ off-shell meson. Again, we expect
to obtain a similar result to the one obtained for $B_s$ off-shell, since this meson has a 
small mass difference when compared with the $B$ meson mass.

We get sum rules for the two structures, $p_\mu$ and $p'_\mu$.
The $p_\mu$ structure does not give good stability in the Borel
mass. Therefore, it is not considered. The other structure, $p'_\mu$, gives
good stability with the pole being bigger than the continuum contribution.
The QCDSR results for the form factor can be extrapolated by 
\begin{equation}
g_{B_s BK^*}^{(B)}(Q^2)= \frac{104.71}{66.97+Q^2} \;
\label{monoBoff}
\end{equation}
 and the coupling constant is
$$
%\begin{equation}
g_{B_s BK^*}= 2.77. 
%\label{couplingbssoff}
%\end{equation} 
$$
This is a very similar result to the one obtained in Table \ref{couplings}.

\section{Conclusion}

We have used the three-point QCD sum rules to study the form factors in the vertices
$B_s B^* K$ and $B_s B K^*$. In each case, we have considered two different sum rules, for two
different mesons off-shell. We have studied the Borel stability and pole dominance in each
case and have determined the Borel window where the sum rules can be used. To get the coupling
constant in each vertex, we have used  the extrapolation method developed in previous works
\cite{Bracco:2011pg}. This extrapolation method has a systematic error which comes from the 
choice of the analytic form of the extrapolating functions. We consider only monopole, 
exponential and Gaussian parametrization:
$
g_{ M M_2 M_3 }^{(I)}(Q^2)= A \exp^{-Q^2/B} 
$, $ g_{ M M_2 M_3 }^{(I)}(Q^2) = A \exp^{-(Q^2/B)^2} $ and 
$ g_{M M_2 M_3 }^{(I)}(Q^2)=\frac{A}{B+Q^2} $.

There are no physical reason for using these forms. However, they are the usual forms used by 
experimentalists and also they have only two parameters, $A$ and $B$, that present some 
regularity, in all our form factors. For instance, when the heavy meson is off-shell, the 
form factor is harder as a monopolar function, and the $B$ cut-off parameter is a big number. 
By the other hand, when the light meson is off-shell, the parametrization is usually 
exponential or Gaussian  and have a smaller $B$ cut-off parameter. 
These cut-offs are showed in Table \ref{cut-off}. 

We have also performed  a very extensive study of the uncertainties to estimate the errors in 
the coupling constants. After considering all good sum rules and including the uncertainties 
study, we obtain the coupling constant  equal to: 
\begin{equation}
g_{B_s B^* K}=8.41 \pm 1.23 \label{finalcoupling:bsbestk}
\end{equation} 
and 
\begin{equation}
g_{B_s BK^*}=3.3 \pm 0.5 \label{finalcoupling:bsbkest}
\end{equation}

\begin{table}[h!]
\begin{center}
\begin{tabular}{ccccc}\hline
    &    \multicolumn{4}{c} {\textbf{cut-off parameters}} \\
		&    \multicolumn{2}{c} {\textbf{$B_s B^* K$}}  & \multicolumn{2}{c} {\textbf{$B_s B K^*$}} \\
  Off-shell meson                    &A & B& A & B \\ \hline 
 $\;\;B_s$                &     $3.18 $    &   $32.98$   & $\;\;80.88$    &   $58.86$\\
	$\;\;B^*$           &  $1.21$  &  $13.65$   &  --      & --\\
 $\;K$       &   $8.25$   & $\;\;6.68$ &   --         & --\\
	$B$          &     --      &   --  &    $104.71$ &$ 66.97 $  \\
 $\;K^* $       &  --   & -- &     $\;\;\;\;2.49$     &$\;\;2.63$     \\\hline
\end{tabular}
\caption{Cut-off parameters for both vertices. }
\label{cut-off}
\end{center}
\end{table}

% --------------B*s off-p'p'

%\begin{figure}[ht]
%\begin{center}
%{\epsfig{figure=polo-conti-Bx-off.eps,height=55mm}}
%\epsfig{figure=estabilidade-Bx-off.eps,height=55mm}
%\epsfig{figure=form-factor-Bs.eps,height=50mm}c
%\caption{(a) The pole-continuum contributions for $B_s B^* K$  and (b) Stability of $g^{(B^*)}_{B_s B^* K}(Q^2=1 %\; GeV^2, M^2)$ for different thresholds parameters.} 
%\label{estabilidadeBss}
%\end{center}
%\end{figure}

We can compare our results with other theoretical predictions using arguments
of heavy hadron chiral perturbation theory (HHChPT), where the
couplings for the bottom-light vertex $g_{ B^*_s B K}$ are related
to the charm-light vertex $g_{D_s D^* K} $ through the relation
\cite{wise,casalbuoni}: 
\beq g_{B_{s} B^* K} = g_{ D^*_s D
K} \frac{m_B}{m_D}, \label{t1} 
\eeq 
where $ m_{B^*} = 5.2$ and $m_D=1.8693$ are the experimental masses. For $g_{D_s D^*
K}$ we can use our previous QCDSR result \cite{angelo06} of $g_{ D^*_s DK}=
3.02$. Therefore, we obtain $ g_{B_{s} B^* K} = 8.39 $, which is in excelent
agreement with our result. This result shows that the relation with charm-light vertex 
can give a good estimate of the couplings in the bottom-light vertex. The uncertainties are 
about 20\%, that is in complete agreement to the technique of QCDSR.
The greatest source of uncertainties found in this work is due to the mass of the bottom quark.

\section*{Acknowledgements}

This work has been partly supported by the Brazilian funding agencies CAPES, CNPq and FAPESP. 

\bibliography{BsBeKBsBKe}

\end{document}